\documentclass[aps,prd,12pt,nofootinbib,preprintnumbers,eqsecnum,showpacs]{revtex4}

\usepackage{bm,amssymb,amsmath,amsfonts,amsbsy,graphicx,simplewick}

\newcommand{\mpl}{m_{\rm Pl}}
\newcommand{\calP}{{\cal P}}
\newcommand{\calR}{{{\cal R}_c}}
\newcommand{\bx}{{\bm x}}
\newcommand{\by}{{\bm y}}
\newcommand{\bp}{{\bm p}}
\newcommand{\bk}{{\bm k}}

\begin{document}

\preprint{YITP-10-82}

\title{Waterfall field in hybrid inflation
and curvature perturbation
}

\author{
Jinn-Ouk Gong$^{1,2}$\footnote{jgong\_AT\_ lorentz.leidenuniv.nl}
and Misao Sasaki$^{2,3,4}$\footnote{misao\_AT\_yukawa.kyoto-u.ac.jp}
}

\affiliation{
~\\
${}^1$Instituut-Lorentz for Theoretical Physics, Universiteit Leiden,
2333 CA Leiden, The Netherlands
\\
${}^2$Yukawa Institute for Theoretical Physics, Kyoto University,
Kyoto 606-8502, Japan
\\
${}^3$Korea Institute for Advanced Study,
Seoul 130-722, Republic of Korea
\\
${}^4$Arnold Sommerfeld Center for Theoretical Physics,
Ludwig-Maximilians-Universit\"at M\"unchen,
80333 M\"unchen, Germany}

\date{\today}

\begin{abstract}
~\\
We study carefully the contribution of the
waterfall field to the curvature perturbation at the end of hybrid inflation.
In particular we clarify the parameter dependence analytically under
reasonable assumptions on the model parameters.
After calculating the mode function of the waterfall field,
we use the $\delta{N}$ formalism and confirm the previously obtained result
that the power spectrum is very blue with the index 4 and is absolutely negligible
on large scales. However, we also find that the resulting curvature perturbation
is highly non-Gaussian and hence we calculate the bispectrum. We find that the
bispectrum is at leading order independent of momentum and exhibits its peak at
the equilateral limit, though it is unobservably small on large scales.
We also present the one-point probability distribution function of the
curvature perturbation.

\end{abstract}


\maketitle

\newpage

\section{Introduction}

Currently, primordial inflation~\cite{inflation} is supposed to be the
leading candidate to provide the necessary conditions for
the successful big bang cosmology~\cite{book}.
The simplest model of inflation driven by only a
single inflaton field is consistent with most recent
observations~\cite{Komatsu:2010fb}. It is however expected that, in the
context of theories beyond the standard model of particle physics e.g.
 supersymmetry, there is a number of multiple scalar fields which may
contribute to the inflationary dynamics~\cite{Lyth:1998xn}.
Furthermore, we may be able to observationally detect deviations from the
predictions of single field models in the near future and to discuss interesting
phenomenology, such as isocurvature perturbations and non-Gaussianity.

Hybrid inflation~\cite{hybrid} is an interesting realization with two
field contents, the usual inflaton field $\phi$ which drives slow-roll
inflation and the waterfall field $\chi$ which terminates
inflation by triggering an instability, a ``waterfall'' phase transition.
Previously, it has been assumed that $\chi$ becomes momentarily massless
only at the time of waterfall and very heavy otherwise, and thus does not
 contribute to the curvature perturbation $\calR$ on large scales: only
the quantum fluctuations of $\phi$ contributes to $\calR$ and we can follow
 the well-known calculations of single field case, with the energy density
 of the universe being dominated by a non-zero vacuum energy.

This naive picture has been receiving a renewed
interest~\cite{Lyth:2010ch,Abolhasani:2010kr,Fonseca:2010nk} with the
common qualitative results that the power spectrum of the curvature perturbation
induced by the waterfall field is very blue and extremely small on large
scales\footnote{For early attempts, see e.g. Ref.~\cite{tpreheating}.}.
However, quantitatively it is not clear if they all agree or not.
In particular, in Ref.~\cite{Fonseca:2010nk}
the $\delta N$ formalism,
which takes account of fluctuations only on super-horizon scales by construction,
was employed to derive the power spectrum,
but the
approach there was not quantitative enough and hence the
dependence on the model parameters was not explicitly presented.

In this note, we provide another complementary view.
We adopt a few reasonable assumptions on the model parameters
and solve the mode functions of $\chi$ in terms
of the number of $e$-folds analytically.
Then using the $\delta{N}$ formalism~\cite{deltaN} we
calculate the corresponding $\calR$ induced by $\chi$ explicitly.

The result is consistent with the above references, i.e.
the contribution of $\chi$ to the
large scale curvature perturbation is totally negligible. We also clarify
the model parameter dependence on the spectrum of the curvature perturbation.
Furthermore, we calculate the corresponding bispectrum, which shows its peak
at the equilateral limit. We also compute explicitly the one-point
probability distribution function which clearly shows the highly non-Gaussian
nature of the curvature perturbation.

The outline of this note is as follows. In Section~\ref{sec:solution},
we find the mode function solution of the waterfall field $\chi$ valid
both on super-horizon and sub-horizon scales. In Section~\ref{section:R},
we calculate the corresponding curvature perturbation $\calR$ induced by
$\chi$ using the $\delta{N}$ formalism. In Section~\ref{sec:correlation},
we present the power spectrum and bispectrum of $\calR$.
In Section~\ref{sec:distribution},
we show the explicit form of the one-point probability distribution function of $\calR$
 and discuss relates issues. We conclude in Section~\ref{sec:conclusion}.
In Appendices, we discuss some technical details.
In Appendix~\ref{app:conventional}, to check the consistency of the $\delta N$ formalism
with the standard perturbation theory, we give an estimation of
the curvature perturbation by using the linear perturbation equation
for $\calR$. We find a good agreement with our result based on
the $\delta N$ formalism.
In Appendix~\ref{app:subhorizon} we reconsider the splitting of the super- and sub-horizon
modes and compute the average over the horizon scales. The results agree with the formulae
we use in the main text.

\section{Mode function solution of waterfall field}
\label{sec:solution}

Before we begin explicit computations, first of all we make the physical
picture clear. Our purpose is to calculate the contribution of the waterfall
field $\chi$ to the curvature perturbation $\calR$. This is only possible when
$\chi$ becomes dynamically relevant. While $\chi$ is well anchored at its minimum
 during the phase of slow-roll inflation and hence does not participate in the
inflationary dynamics, it controls the physical processes from the moment of
waterfall till the end of inflation. Thus, in the context of the $\delta{N}$ formalism,
 if we can find the evolution of $\chi$ during this phase as a function of the
number of $e$-folds $N$, it amounts to finding $\calR$ by the geometrical
identity $\calR = \delta{N}$. Therefore, our aim in this section is to calculate
 $\chi = \chi(N)$ starting from the moment of waterfall.
 We will directly use this result to calculate $\calR$ in the next section.

We consider the potential of the two fields, the inflaton $\phi$ and
 the waterfall field $\chi$, as
\begin{equation}
V(\phi,\chi) = \frac{\lambda}{4} \left( \frac{M^2}{\lambda} - \chi^2 \right)^2 + \frac{1}{2}m^2\phi^2 + \frac{1}{2}g^2\phi^2\chi^2 \, .
\end{equation}
We note that during the most period of inflation of our interest,
it is assumed that the vacuum energy $V_0 = M^4/(4\lambda)$ dominates so that the
Hubble parameter is effectively a constant, $H=H_0$.
This is a good approximation even after the waterfall phase transition
until the last moment of inflation. The slow-roll and the waterfall conditions
are
\begin{align}
\frac{m^2}{H_0^2} \ll & 1 \, ,
\\
\frac{M^2}{H_0^2} \equiv \beta \gg & 1 \, ,
\end{align}
respectively.

The equations of motion are given by
\begin{align}
\label{phieq}
\ddot\phi + 3H\dot\phi + \left( m^2 + g^2\chi^2 \right)\phi = & 0 \, ,
\\
\label{chieq}
\ddot\chi + 3H\dot\chi - \frac{1}{a^2}\nabla^2\chi
 + \left( -M^2 + g^2\phi^2 + \lambda\chi^2 \right)\chi = & 0\,,
\end{align}
where the spatial gradient term for $\phi$ is neglected as usual.
Note that before waterfall, $\phi^2>\phi_c^2\equiv M^2/g^2$,
$\chi$ is well anchored at its minimum $\chi=0$ so it is itself
the same as its fluctuation, $\chi = \delta\chi$.
Thus we may regard (\ref{chieq}) as the equation for $\delta\chi$\footnote{Note that
during inflation $\delta\rho_\phi \sim \delta\phi$ while
$\delta\rho_\chi \sim \delta\chi^2$, and thus the metric fluctuations
 are {\em relatively} second order with respect to $\delta\chi$ and
does not appear in the equation of motion for $\delta\chi$. This
situation is closely analogous to the case of false vacuum
inflation~\cite{Gong:2008ni}. The correlation functions also show similar
momentum dependence to those produced during false vacuum inflation~\cite{Gong:2008ni,Gong:2009dh}},
which arises from the vacuum fluctuations. Then after the waterfall
transition, $\delta\chi$ becomes unstable and $\delta\chi^2$ starts to
grow rapidly, and inflation ends when the inflaton starts to roll
fast, which happens when the term $g^2\delta\chi^2$
exceeds $m^2$ in (\ref{phieq}). Here we adopt the mean field approximation,
i.e. we replace $g^2\delta\chi^2$
by its expectation value $g^2\langle\delta\chi^2\rangle$,
which should be valid for the motion of the homogeneous inflaton field $\phi$.
We also assume that the nonlinear term $\lambda\delta\chi^2$ in (\ref{chieq})
can be neglected until the end of inflation. That is,
we assume
\begin{eqnarray}
M^2\gg\frac{\lambda}{g^2}m^2\gtrsim\lambda\langle\delta\chi^2\rangle\,.
\label{linearchi}
\end{eqnarray}
At the end of calculation, we must check if this condition is satisfied
for the range of the parameters of our interest.

We can rewrite (\ref{phieq}) and (\ref{chieq}) in a more convenient form by
using the number of $e$-folds as the time variable, $dN=Hdt$.
Denoting the derivative with respect to $N$ by a prime, 
we write
\begin{align}
\label{phieq2}
\phi'' + 3\phi' + \left( \frac{m^2}{H_0^2}
+ g^2\frac{\langle\delta\chi^2\rangle}{H_0^2} \right) \phi = & 0 \, ,
\\
\label{chieq2}
\delta\chi'' + 3\delta\chi' - \frac{1}{a^2H_0^2}\nabla^2\delta\chi
 + \left( -\beta + g^2\frac{\phi^2}{H_0^2}
+ \lambda\frac{\delta\chi^2}{H_0^2} \right) \delta\chi = & 0 \, .
\end{align}
Let $N_c$ be the time at which the waterfall transition occurs, $\phi(N_c)=\phi_c=M/g$.
Before waterfall, since $\delta\chi$ is very massive, $g^2\phi^2\gg H^2$,
it is dominated by the standard vacuum fluctuations and the bare
expectation value $\langle\delta\chi^2\rangle$ is ultraviolet divergent.
Here we regularize it so that it vanishes before waterfall,
$\langle\delta\chi^2\rangle=0$ at $N<N_c$.
Then (\ref{phieq2}) is easily solved to give
\begin{equation}\label{phisol}
\phi = \phi_c e^{-rn} \, ,
\end{equation}
where $n=N-N_c$ is the number of $e$-folds measured relative to the time
of the waterfall transition, and we have introduced the parameter $r$ by
\begin{eqnarray}
r \equiv \frac{3}{2} - \sqrt{\frac{9}{4} - \frac{m^2}{H_0^2}}
\approx \frac{m^2}{3H_0^2} \ll 1 \, .
\end{eqnarray}
We note that we can write the scale factor $a$ and the conformal time
$\eta=-1/(aH)$ using $n$ as
\begin{align}
a = & a_ce^n = \frac{k_c}{H_0}e^n \, ,
\\
\label{conformal_time}
\eta = &- \frac{e^{-n}}{a_cH_0} = -\frac{e^{-n}}{k_c} \, ,
\end{align}
respectively, where $a_c=a(N_c)$ and $k_c = a_cH_0$.

Inserting the background solution (\ref{phisol}) for $\phi$
into (\ref{chieq2}) and neglecting the nonlinear term
in accordance with the assumption (\ref{linearchi}),
we obtain the equation for $\delta\chi$ in the Fourier space,
\begin{eqnarray}
\label{linchieq}
\delta\chi_{\bm k}'' + 3\delta\chi_{\bm k}' +
 \left[\frac{k^2}{k_c^2}e^{-2n}+\beta \left( e^{-2rn} - 1 \right)\right]
 \delta\chi_{\bm k} = 0 \,.
\end{eqnarray}

\subsection{High frequency limit $k/a\to\infty$: WKB solution}

In the high frequency limit, we can solve (\ref{linchieq}) in terms
of the WKB approximation. In this limit the proper asymptotic behavior
of the positive frequency function is given by
\begin{eqnarray}
\delta\chi_{k} \underset{k\to\infty}{\longrightarrow} \frac{e^{-ik\eta}}{\sqrt{2k}a}=
\frac{H_0}{\sqrt{2k_c^3}\sqrt{k/k_c}}\exp\left(-i\frac{k}{k_c}\int^ndne^{-n}\right)\,.
\end{eqnarray}
The WKB solution that has this asymptotic behavior is readily obtained
as
\begin{eqnarray}
\label{WKBsol}
\delta\chi_{k} = \frac{e^{-n}H_0}{\sqrt{2k_c^3}
 \left[ \left(k/k_c\right)^2 + \tilde{\beta} e^{2n} \right]^{1/4}}
 \exp \left[ -i \int^ndne^{-n}
\sqrt{ \left( \frac{k}{k_c} \right)^2 + \tilde{\beta} e^{2n} } \right] \, ,
\end{eqnarray}
where for convenience we have defined $\tilde\beta$ by
\begin{eqnarray}
\tilde\beta\equiv\beta(e^{-2rn}-1)\,.
\label{tildebeta}
\end{eqnarray}
The above WKB solution is valid for any $k$ at sufficiently
early times, $-n\gg1$.

\subsection{Low frequency limit $k/a\to0$: Hankel function solution}

In the large scale limit $k\to0$, (\ref{linchieq}) becomes
\begin{equation}\label{chieq3}
\delta\chi_0'' + 3\delta\chi_0' + \beta \left( e^{-2rn} - 1 \right) \delta\chi_0 = 0 \, .
\end{equation}
Then, the solution is easily found to be
\begin{equation}\label{LSsolution}
\delta\chi_0(n) = e^{-3n/2} \left[ c_1 H_\nu^{(1)}
\left( \frac{\sqrt{\beta}}{r}e^{-rn} \right) + c_2 H_\nu^{(2)}
\left( \frac{\sqrt{\beta}}{r}e^{-rn} \right) \right] \, ,
\end{equation}
where $H_\nu^{(1)}$ and $H_\nu^{(2)}$ are the Hankel function of first
and second kind, respectively, and are complex conjugate to each other,
$c_1$ and $c_2$ are constants to be determined, and
\begin{equation}
\nu \equiv \frac{\sqrt{\beta + 9/4}}{r}\approx\frac{\sqrt{\beta}}{r} \, .
\end{equation}

\subsection{Large scale modes: $k\ll k_c$}
\label{largescalemodes}

Now let us consider the long wavelength modes $k\ll k_c$ which
are already on super-horizon scales by the time of the waterfall transition.
For these modes, we match the WKB solution to the Hankel function
solution at some time well before the waterfall, $n<0$ and $|n|\gg 1$.

In the limit $k/k_c\to0$, the WKB solution (\ref{WKBsol})
becomes
\begin{equation}\label{WKB_asymptotic}
\delta\chi_{k} \underset{k\to0}{\longrightarrow}
\frac{e^{-3n/2}H_0}{\sqrt{2rk_c^3}}
\left( \frac{\sqrt{\beta}}{r}e^{-rn} \right)^{-1/2}
\exp \left( i\frac{\sqrt{\beta}}{r}e^{-rn} \right) \, ,
\end{equation}
where we have assumed $e^{-2rn}\gg1$.

As for the Hankel function solution, the argument
is very large in the limit $-n\gg 1$,
 $\sqrt{\beta}e^{-rn}/r\approx\nu e^{-rn}\gg\nu$.
Thus using the asymptotic form of the Hankel function,
\begin{equation}
H_\nu^{(1)}(z) \underset{z\gg\nu}{\longrightarrow} \sqrt{\frac{2}{\pi z}}
 \exp \left[ i \left( z - \frac{\nu}{2}\pi - \frac{\pi}{4} \right) \right] \, ,
\end{equation}
we find that (\ref{LSsolution}) becomes
\begin{equation}\label{LSsolution_limit}
\delta\chi_k \underset{-n\gg1}{\longrightarrow} c_1 e^{-3n/2}
 \sqrt{\frac{2}{\pi}}\left( \frac{\sqrt{\beta}}{r}e^{-rn} \right)^{-1/2}
 e^{-i(\nu\pi/2+\pi/4)} \exp \left( i\frac{\sqrt{\beta}}{r}e^{-rn} \right) + \cdots \, ,
\end{equation}
where for notational simplicity we have omitted the term
proportional to $H_\nu^{(2)}$ whose coefficient is $c_2$.

Comparing (\ref{LSsolution_limit}) with (\ref{WKB_asymptotic}), we
see that $H_\nu^{(1)}$ gives the correct phase factor dependence
of (\ref{WKB_asymptotic}) and thus we have $c_2 = 0$ and
\begin{equation}
c_1 = \sqrt{\frac{\pi}{2}} \frac{H_0}{\sqrt{2rk_c^3}} e^{i(\nu\pi/2+\pi/4)} \, .
\end{equation}
Thus, the long wavelength positive frequency function is given by
\begin{equation}\label{general_sol}
\delta\chi_{k} \underset{k\ll k_c}{\longrightarrow}
 e^{-3n/2} \sqrt{\frac{\pi}{2}}
\frac{H_0}{\sqrt{2rk_c^3}} e^{i(\nu\pi/2+\pi/4)}
H_\nu^{(1)} \left( \frac{\sqrt{\beta}}{r}e^{-rn} \right) \, .
\end{equation}

Here let us evaluate the mode function
at the moment of waterfall $n=0$.
At $n=0$, remembering that $\beta\gg1$, the Hankel function takes the form
\begin{equation}
H_{\nu}^{(1)} \left( \frac{\sqrt{\beta}}{r} \right)
 \approx H_{\nu}^{(1)} \left(\nu\right) \, ,
\end{equation}
with $\nu \approx \sqrt{\beta}/r$.
That is, the index and the argument of the Hankel function are the same.
 In this case, the Hankel function solution takes the form
\begin{equation}
H_\nu^{(1)}(\nu) =
 \left( \frac{6}{\nu} \right)^{1/3} \frac{2}{3\Gamma(2/3)} e^{-i\pi/3} \, .
\end{equation}
Then, denoting by a subscript $L$ the long wavelength modes
which are on super-horizon scales at $n=0$, we can write
\begin{equation}\label{super_initial}
\delta\chi_L(n=0) \underset{k\to0}{\longrightarrow}
\frac{2\sqrt{\pi}}{3^{2/3}\Gamma(2/3)} \frac{H_0}{\sqrt{2k_c^3}\alpha^{1/3}}
 \exp \left[ i \left( \nu - \frac{1}{6} \right) \frac{\pi}{2} \right] \, ,
\end{equation}
where the numerical factor reads
$2\sqrt{\pi}/\left[3^{2/3}\Gamma(2/3)\right] \approx 1.25854$,
and we have defined\footnote{
Our $\alpha$ is equal to $\epsilon_\psi$ in Ref.~\cite{Abolhasani:2010kr}. }
\begin{equation}
\alpha \equiv \sqrt{2r\beta} \, .
\end{equation}
As we will see in the next section, we must require $\alpha\gg1$.
The above result (\ref{super_initial}) implies that
all the super-horizon modes have the same amplitude
at the moment of waterfall. The the moment of waterfall will be
taken as the ``initial'' time to estimate the contribution of
$\delta\chi$ to the curvature perturbation $\calR$.

\subsection{Small scale modes: $k\gg k_c$}

For the modes that are still on sub-horizon scales at the time of waterfall,
$k\gg k_c$, the WKB solution is valid until $n=0$.
Denoting them by a subscript $S$, (\ref{WKBsol}) readily gives
\begin{equation}
\delta\chi_S = \frac{H_0}{\sqrt{2k}k_c}e^{-n}
 \exp \left( i\frac{k}{k_c}e^{-n} \right) \, ,
\end{equation}
so that at the moment of waterfall
\begin{equation}\label{sub_initial}
\delta\chi_S(n=0) = \frac{H_0}{\sqrt{2k}k_c} e^{ik/k_c} \, .
\end{equation}
This is the ``initial'' amplitude of the sub-horizon modes.

\vspace{10mm}
Before we move on, we mention that the initial amplitudes of large
scale limit (\ref{super_initial}) and that of small scale
limit (\ref{sub_initial}) do not match at $k=k_c$ if we extrapolate
from both sides, but are different by a suppression factor $\alpha^{-1/3}$.
This indicates that in the intermediate regime around $k=k_c$ these
two extreme values are deviating from the limiting values
and smoothly connected~\cite{Fonseca:2010nk}.
In particular, this implies that the sub-horizon modes with $k\gtrsim k_c$
have slightly different initial amplitudes from (\ref{sub_initial}).
However this will not affect our subsequent discussions
because of the phase volume $\sim k^3$ that gives rise to a sharp peak
in the spectrum at $k\approx\alpha k_c\gg k_c$, as we will see below.
Hence we just use (\ref{super_initial}) for the initial amplitude
of the large scale modes with $k<k_c$
and (\ref{sub_initial}) for that of the small scale modes with $k>k_c$.

\subsection{Evolution of the relevant modes after waterfall}
\label{evolution}

Having found the ``initial'' amplitudes of both large scale and
small scale modes, now we can calculate the subsequent evolution of the modes
until the end of inflation.

Let us first consider the large scale modes.
The solution is given by (\ref{general_sol}) and
is valid for $n > 0$ as well. Then, using the asymptotic form of
 the large $\nu = \sqrt{\beta}/r$ we can find~\cite{mathbook}
\begin{equation}\label{Hankel_asymptotic}
H_{\nu}^{(1)}\left( \nu e^{-rn} \right)
 = \sqrt{\frac{2r}{\pi\alpha}}
 \exp \left( \frac{2}{3}\alpha n^{3/2} - \frac{1}{4}\log{n} \right)
e^{-i\pi/2} \, .
\end{equation}
This is obtained with $rn\ll1$, and is thus valid for $n\ll1/r$.
For any sensible model of hybrid inflation $r\ll1$, while the number
of $e$-folds after waterfall until the end of inflation, $n_f$,
is $\mathcal{O}(1)$ or at most a few.
Hence this asymptotic form is valid until the end of inflation.
Then, plugging (\ref{Hankel_asymptotic}) into (\ref{general_sol}),
 we can find that after waterfall the mode function on super-horizon
 scales evolves as\footnote{Note that the same dependence on the number
 of $e$-folds was found from the Airy function solutions in
Refs.~\cite{Lyth:2010ch,Abolhasani:2010kr}. But the corresponding
equation solved in these references is a particular limit of the
general equation (\ref{chieq3}) and thus so does the solution,
 as we show here explicitly.}
\begin{align}\label{super_after}
|\delta\chi_L(n)| = & \frac{H_0}{\sqrt{2\alpha k_c^3}}
 \exp \left( \frac{2}{3}\alpha n^{3/2} - \frac{3}{2}n - \frac{1}{4}\log{n} \right)
\nonumber\\
= & \left| \delta\chi_L(n=0) \right| \frac{3^{2/3}
\Gamma(2/3)}{2\sqrt{\pi}} \alpha^{-1/6}
 \exp \left( \frac{2}{3}\alpha n^{3/2} - \frac{3}{2}n - \frac{1}{4}\log{n} \right) \, ,
\end{align}
where the initial amplitude of the large scale modes $\delta\chi_L(n=0)$
 is given by (\ref{super_initial}). As the logarithmic term indicates,
 (\ref{super_after}) does not hold precisely at $n=0$ but is valid for,
 as mentioned above, some time after waterfall till the end of inflation.
 As we will evaluate $\delta\chi$ at the end of inflation
 $n_f=\mathcal{O}(1)$, we can justifiably use (\ref{super_after})
 to calculate the curvature perturbation.

Let us now turn to the small scale modes.
An important point to calculate the evolution of sub-horizon modes is that
 the end of inflation is determined by the quanta of $\chi$ which become
 tachyonic right after waterfall~\cite{Abolhasani:2010kr}, and affect the
effective mass of $\phi$ in the form $g^2\langle\delta\chi^2\rangle$.
 The modes which become tachyonic satisfy, by definition,
$(k/k_c)^2<|\tilde\beta|$ in (\ref{linchieq}).
Assuming $n=\mathcal{O}(1)$, we have $|\tilde\beta|\approx2\beta rn\sim \alpha^2$.
Hence we find that the modes with
\begin{equation}
\frac{k}{k_c} \lesssim \alpha
\end{equation}
become tachyonic. Thus $\alpha$ must be much greater than unity in order
to have an effective tachyonic instability.

To summarize, the small scale modes of our interest, which contribute
to the tachyonic instability and control the end of inflation,
are those in the interval
\begin{eqnarray}
k_c\lesssim k\lesssim\alpha k_c\,.
\end{eqnarray}
Since $(k/k_c)^2$ can be neglected
in comparison with $\tilde\beta$ at leading order approximation,
they satisfy the same equation as the
equation for the large scale modes, (\ref{chieq3}).
Hence the evolution of these modes at $n>0$ is the same as
that given by (\ref{super_after}).
That is,
\begin{equation}\label{sub_after}
\left| \delta\chi_S(n) \right| = \left| \delta\chi_S(n=0) \right|
A\,\exp
\left( \frac{2}{3}\alpha n^{3/2} - \frac{3}{2}n - \frac{1}{4}\log{n} \right) \, ,
\end{equation}
where we have set the overall coefficient as
\begin{equation}
A\equiv \frac{3^{2/3}\Gamma(2/3)}{2\sqrt{\pi}}\alpha^{-1/6}\,.
\end{equation}

\section{Curvature perturbation induced by waterfall field}
\label{section:R}

In this section, we calculate the curvature perturbation $\calR$
by using (\ref{super_after}) and (\ref{sub_after}) in the context
 of the $\delta{N}$ formalism. In the $\delta N$ formalism
the spacetime geometry is spatially smoothly varying over super-horizon
scales while each Hubble horizon size region is regarded as a homogeneous and
isotropic universe. Hence we first need to smooth over the horizon scale
$H_0^{-1}$,
\begin{align}\label{deltapsisquare}
\delta\chi^2(n)= &
\left[ \delta\chi_L^2(0)
+ \left\langle \delta\chi_S^2(0) \right\rangle \right] A^2
 \exp \left( \frac{4}{3}\alpha n^{3/2} - 3n \right)
\nonumber\\
= & \left[ \delta\chi_L^2(0) + \int_{k_c}^{\alpha k_c}
\frac{d^3k}{(2\pi)^3} \delta\chi_S^2(0)
\right] A^2 \exp \left( \frac{4}{3}\alpha n^{3/2} - 3n \right) \, ,
\end{align}
where $\delta\chi_L^2(0)$ and hence $\delta\chi^2(n)$ is spatially varying
on super-horizon scales.
Note that we have omitted the logarithmic dependence term on $n$
in the exponent, which is sub-dominant when we evaluate at
 $n=n_f=\mathcal{O}(1)$. We have also subtracted the contribution from the modes
with $k>\alpha k$ since they remain stable and behave in the same way
as the flat Minkowski vacuum modes, in accordance with the regularization
we adopted, i.e. $\langle\delta\chi^2(n)\rangle=0$ at $n<0$.

With $\alpha\gg1$, from the initial
amplitudes (\ref{super_initial}) and (\ref{sub_initial}) we can
 see that the contribution of sub-horizon modes is much bigger than the one from super-horizon modes
if the average is taken. At the end of inflation we have
$\langle \delta\chi^2(n_f)\rangle=m^2/g^2$,
 so that using
\begin{equation}\label{shortdeltapsisquared}
\left\langle \delta\chi_S^2(0)  \right\rangle
= \frac{\alpha^2H_0^2}{8\pi^2} \, ,
\end{equation}
which follows from (\ref{sub_initial}), we have
\begin{eqnarray}
\frac{m^2}{g^2}
=\langle \delta\chi^2(n_f)\rangle
&=&\left[\langle\delta\chi_L^2(0)\rangle
+\langle\delta\chi_S^2(0)\rangle\right]
A^2\exp \left( \frac{4}{3}\alpha n_f^{3/2} - 3n_f \right)
\cr
&\approx&
\frac{\alpha^2H_0^2}{8\pi^2}
A^2\exp \left( \frac{4}{3}\alpha n_f^{3/2} - 3n_f \right)\,,
\end{eqnarray}
 we find
\begin{equation}\label{nf}
\exp \left( \frac{4}{3}\alpha n_f^{3/2} - 3n_f \right)
 = \frac{8\pi^2m^2}{g^2\alpha^2A^2H_0^2} \, .
\end{equation}

Now let us rephrase the above discussion in a form more convenient
 for the $\delta{N}$ formalism. With coordinate dependence
 explicit, $\delta\chi^2$ given by
 (\ref{deltapsisquare}) is recast as
\begin{equation}
\delta\chi^2(n,{\bm x})
= \delta\chi_L^2(n,{\bm x})
+ \left\langle \delta\chi_S^2(n) \right\rangle \, .
\end{equation}
As mentioned in the first paragraph of this section,
since the smoothing is done over the horizon scales $H_0^{-1}$,
there remains no spatial coordinate dependence in
 $\left\langle \delta\chi^2_S \right\rangle$.
Meanwhile, we do have a spatial coordinate
dependence for the modes with wavelengths longer than $H_0^{-1}$,
 which is what we should take care of in the context
of the $\delta{N}$ formalism.
Neglecting $-3n$ in the exponential for simplicity since $\alpha \gg 1$,
splitting $n = \bar{n} + \delta{n}$ and expanding in terms
of $\delta{n}$, (\ref{deltapsisquare}) is written as
\begin{equation}
\delta\chi^2(\bar n+\delta n) = \left[ 1 +
\frac{\delta\chi_L^2(0)}{\left\langle \delta\chi_S^2(0) \right\rangle}
 \right]
 \left\langle \delta\chi^2_S(0) \right\rangle
A^2 \exp \left( \frac{4}{3}\alpha\bar{n}^{3/2} \right)
 \left( 1 + 2\alpha\bar{n}^{1/2}\delta{n} + \cdots \right) \, ,
\end{equation}
where $\left\langle \delta\chi^2_S \right\rangle \gg \delta\chi_L^2$
 as discussed above.

Now we evaluate $\delta{n}$ at a later time, say,
at the end of inflation $n = n_f$. Here it is important to note that
the end of inflation is controlled by the value of $\delta\chi^2$
at each spatial point, namely,
\begin{equation}
\delta\chi^2(n_f,{\bm x})= \frac{m^2}{g^2} =
\left\langle \delta\chi^2(n_f) \right\rangle\,.
\label{endofinf}
\end{equation}
Analogous to the case when the value of the inflaton field determines
the end of inflation hypersurface, this condition determines the
end of inflation hypersurface on which the energy density is uniform
(at leading order approximation where the contribution of the inflaton
to the energy density is negligible).
Then, using (\ref{shortdeltapsisquared}) and
(\ref{nf}), we find
\begin{equation}\label{deltapsi_deltan2}
1 \approx \left[ 1 +
 \frac{\delta\chi_L^2(0)}{\left\langle \delta\chi_S^2(0) \right\rangle} \right]
\left( 1 + 2\alpha n_f^{1/2}\delta{n} \right) \, ,
\end{equation}
where we have truncated at linear order in $\delta{n}$.
Inverting this relation, we can write the curvature perturbation
 generated between the moment of phase transition and the end of
 inflation as
\begin{equation}\label{curvpert}
\calR({\bm x}) = \delta{n} \approx -\frac{1}{2\alpha n_f^{1/2}}
\frac{\delta\chi_L^2(0,{\bm x})}{\left\langle \delta\chi_S^2(0) \right\rangle}\, .
\end{equation}
This explicitly shows that the spectrum of $\calR$ is determined by
the spectrum of $\delta\chi_L^2$.

 From the result obtained previously
in Sec.~\ref{largescalemodes}, the mode function is $k$-independent for $k<k_c$.
This implies that the power spectrum of $\delta\chi_L$ is white:
$P_{\delta\chi_L}(k)$ is constant
(in the conventional terminology used in cosmology, it is blue with
the spectral index of $4$:
${\cal P}_{\delta\chi_L}(k)\equiv k^3/(2\pi^2) P_{\delta\chi_L}(k)\propto k^{n-1}$ with $n=4$.
See (\ref{chifluc_spectrum}) for example).
Assuming that the spectrum has a ultraviolet cutoff at $k=k_c$, this implies
that $\delta\chi_L^2$ also has the same white spectrum, since the
convolution of two white spectra is white.
Thus apart from the amplitude which we will calculate below,
we can already conclude that $\calP_{\cal R}\propto k^3$, so that the spectral
 index is strongly blue with $n_\calR = 4$, indicating that the curvature
perturbation is strongly suppressed on large scales.

Before we move to the computation of the power spectrum, let us
 also observe that $\calR$ seems to be {\em always} negative.
This can be also read
from (\ref{deltapsi_deltan2}): although $\delta\chi_L$
may be positive or negative, it appears in the form of a square in
(\ref{deltapsi_deltan2}). So irrespective of
 the sign of $\delta\chi_L$ its contribution is always positive.
Meanwhile, the left hand side of (\ref{deltapsi_deltan2}) is a constant.
 Thus, to compensate the positive contribution of $\delta\chi_L^2$
to make the left hand side a constant, $\delta{n}$ is always negative.
 Also we note that the average value of $\calR$ is {\em not} zero,
\begin{eqnarray}
\langle\calR\rangle=-\frac{1}{2\alpha n_f^{1/2}}
\frac{\langle\delta\chi_L^2(0)\rangle}
{\langle \delta\chi_S^2(0)\rangle}\,.
\end{eqnarray}
We will consider these issues a little further later.

Finally, before closing this section,
let us discuss constraints on the model parameters.
First we consider the condition
 that comes from the fact that
the initial value of $\delta\chi^2$ must be smaller
than the final value of it.
 From (\ref{shortdeltapsisquared}) and (\ref{endofinf}),
we find
\begin{eqnarray}
g^2\ll \frac{24\pi^2r}{\alpha^2}=\frac{12\pi^2}{\beta}\,.
\label{paracond}
\end{eqnarray}
On the other hand, for this hybrid inflation model
to be viable, the amplitude of the curvature perturbation due
to the inflaton field $\phi$ must not exceed the observed value,
$\calP_{\cal R}^{(\phi)}\lesssim 10^{-9}$,
\begin{eqnarray}
10^{-9}\gtrsim \calP_{\cal R}^{(\phi)}
=\left(\frac{H_0^2}{2\pi\dot\phi}\right)_{t_k}^2
>\left(\frac{3H_0^3g}{2\pi m^2M}\right)^2
=\frac{g^2}{(2\pi)^2r^2\beta}\,,
\end{eqnarray}
hence
\begin{eqnarray}
g^2\lesssim(2\pi)^210^{-9}r^2\beta\,.
\label{infcurv}
\end{eqnarray}
We see that both (\ref{paracond}) and (\ref{infcurv}) can be safely
satisfied for reasonable values of the parameters.
As a typical example, consider the case $r=m^2/(3H_0^2)=1/10$
and $\beta=M^2/H_0^2=100$, which implies $\alpha^2=20$.
In this case (\ref{paracond}) gives $g^2\ll1$
while we have $g^2\lesssim 4\times10^{-8}$ from (\ref{infcurv}).
Thus the condition (\ref{paracond}) is well satisfied in
this typical case.

Let us also consider the other conditions on
the model parameters. The requirement (\ref{linearchi})
that the linear approximation
to the equation of motion for $\chi$ is valid
implies the condition on $\lambda$ as, using (\ref{infcurv}),
\begin{eqnarray}
\lambda\ll \frac{g^2M^2}{m^2}
\lesssim \frac{(2\pi)^2}{3}10^{-9}r\,\beta^2\,.
\label{lambdacond}
\end{eqnarray}
This gives $\lambda\ll10^{-5}$ for $r=1/10$ and $\beta=100$.
Another condition of $\lambda$ comes from the observational
constraint on the amplitude of
tensor perturbations, $H^2/\mpl^2\lesssim 10^{-10}$.
In the present model, since
$H^2/\mpl^2=V_0/(3\mpl^4)=M^4/(12\lambda\mpl^4)$,
this gives the condition
\begin{eqnarray}
\frac{M^4}{\mpl^4}\lesssim 10^{-9}\lambda\,.
\end{eqnarray}
On the other hand, from $\beta=M^2/H_0^2$ we have
${M^2}/{\mpl^2}={12\lambda}/{\beta}$.
Therefore we must have
\begin{eqnarray}
\lambda\lesssim 10^{-11}\beta^2\,.
\end{eqnarray}
Comparing with (\ref{lambdacond}), we see that this condition is
also well satisfied for typical values of the model parameters.

\section{Correlation functions}
\label{sec:correlation}

\subsection{Power spectrum}

In this section, we drop the subscript $L$ from $\delta\chi_L$ for
notational simplicity. Since $\langle{\calR}\rangle\neq0$,
it is more relevant to consider ${\calR}-\langle\calR\rangle$
rather than ${\calR}$ itself given by (\ref{curvpert}).
Nevertheless, the difference becomes irrelevant in the Fourier space
as long as we focus on a finite wavenumber.
We will discuss this point in the next section.

Moving to the Fourier space, we can write
\begin{equation}
\calR(\bm k) = -\frac{1}{2\alpha n_f^{1/2}}
 \frac{\left( \delta\chi^2 \right)_{\bm k}}
{\left\langle \delta\chi_S^2 \right\rangle} \, ,
\end{equation}
so that the power spectrum is written as
\begin{align}
\left\langle \calR(\bm k)\calR(\bm q) \right\rangle
\equiv & (2\pi)^3 \delta^{(3)}({\bm k}+{\bm q}) P_{\cal R}(k)
\nonumber\\
= & \frac{1}{4\alpha^2n_f} \frac{\left\langle
 \left( \delta\chi^2 \right)_{\bm k} \left( \delta\chi^2 \right)_{\bm q}
 \right\rangle}{\left\langle \delta\chi_S^2 \right\rangle^2 }
= \frac{16\pi^4}{\alpha^6n_fH_0^4} \left\langle
\left( \delta\chi^2 \right)_{\bm k} \left( \delta\chi^2 \right)_{\bm q}
\right\rangle\,,
\label{Rtochi2}
\end{align}
where we have used (\ref{shortdeltapsisquared}) in the last equality.

Before waterfall, $\delta\chi$ is purely quantum
and it can be expressed in terms of the
creation and annihilation operators $a_{\bm k}^\dag$ and $a_{\bm k}$ as
\begin{equation}\label{Fourier_decomp}
\delta\chi = \int \frac{d^3k}{(2\pi)^3} e^{i{\bm k}\cdot{\bm x}}
\delta\chi_{\bm k} = \int \frac{d^3k}{(2\pi)^3} e^{i{\bm k}
\cdot{\bm x}} \left( a_{\bm k}\chi_k + a_{-{\bm k}}^\dag
\chi_k^* \right) \, ,
\end{equation}
where $a_{\bm k}^\dag$ and $a_{\bm k}$ satisfy the canonical
commutation relations
\begin{equation}
\left[ a_{\bm k}, a_{\bm q}^\dag \right]
= (2\pi)^3\delta^{(3)}({\bm k}-{\bm q}) \, ,
\end{equation}
otherwise zero, and the mode function $\chi_k$ follows the same equation as that of $\delta\chi$. Since the Fourier component of $\delta\chi^2$ is
written as a convolution
\begin{equation}
\left( \delta\chi^2 \right)_{\bm k}
= \int \frac{d^3q}{(2\pi)^3} \delta\chi_{\bm q}
 \delta\chi_{{\bm k}-{\bm q}} \, ,
\end{equation}
we have to correlate four creation and annihilation operators
with different momenta,
\begin{equation}
\left\langle \left( \delta\chi^2 \right)_{\bm k}
\left( \delta\chi^2 \right)_{\bm q} \right\rangle =
 \int \frac{d^3pd^3l}{(2\pi)^{3\cdot2}} \left\langle
\left( \delta\chi_{\bm p}\delta\chi_{{\bm k}-{\bm p}} \right)
\left( \delta\chi_{\bm l}\delta\chi_{{\bm q}-{\bm l}} \right)
\right\rangle \, .
\end{equation}

To calculate the above, we should note that what we are interested
in are connected graphs, correlating {\em different} $(\delta\chi^2)_{\bm k}$'s.
Thus the meaningful contractions are
\begin{equation}
\left\langle \left( \delta\chi^2 \right)_{\bm k}
\left( \delta\chi^2 \right)_{\bm q} \right\rangle =
 \Big\langle
 \contraction[1.5ex]
  {\delta} {\chi_{\bm p}} {\delta\chi_{{\bm k}-{\bm p}})
(\delta\chi} {{}_{\bm l}\delta\chi_{{\bm q}-{\bm l}}}
 \contraction[2.5ex]
  {\delta\chi_{\bm p}\delta} {\chi_{{\bm k}}} {{}_{-{\bm p}})
(\delta\chi} {{}_{\bm l}}
  \left( \delta\chi_{\bm p} \delta\chi_{{\bm k}-{\bm p}} \right)
  \left( \delta\chi_{\bm l} \delta\chi_{{\bm q}-{\bm l}} \right)
 \Big\rangle +
 \Big\langle
 \contraction[1.5ex]
  {\delta} {\chi_{\bm p}} {\delta\chi_{{\bm k}-{\bm p}})(\delta\chi} {{}}
 \contraction[2ex]
  {\delta\chi_{\bm p}} {\delta\chi_{{\bm k}-}} {{}_{{\bm p}})(\delta\chi_{\bm l}\delta} {\chi_{\bm q}}
  \left( \delta\chi_{\bm p} \delta\chi_{{\bm k}-{\bm p}} \right)
  \left( \delta\chi_{\bm l} \delta\chi_{{\bm q}-{\bm l}} \right)
 \Big\rangle \, ,
\end{equation}
while the remaining possible contractions are within the
 {\em same} $(\delta\chi^2)_{\bm k}$'s and hence are irrelevant.
Then, we can easily find
\begin{align}
\left\langle \left( \delta\chi^2 \right)_{\bm k}
 \left( \delta\chi^2 \right)_{\bm q} \right\rangle
= & \chi_p\chi_{|{\bm k}-{\bm p}|}
\chi_l^*\chi_{|{\bm q}-{\bm l}|}^* (2\pi)^{3\cdot2}
\nonumber\\
& \times \left[ \delta^{(3)}({\bm p}+{\bm q}-{\bm l})
\delta^{(3)}({\bm k}-{\bm p}+{\bm l})
+ \delta^{(3)}({\bm p}+{\bm l})\delta^{(3)}
({\bm k}-{\bm p}+{\bm q}-{\bm l}) \right] \, .
\end{align}
Thus, eliminating one of the momenta using the delta functions,
 and using the remaining delta function
 $\delta^{(3)}({\bm k}+{\bm q})$ to
replace ${\bm q}$ with $-{\bm k}$, we find
\begin{equation}\label{mode_integration}
\left\langle \left( \delta\chi^2 \right)_{\bm k}
\left( \delta\chi^2 \right)_{\bm q} \right\rangle
= 2 \int d^3p |\chi_p|^2\left|\chi_{|{\bm k}-{\bm p}|}\right|^2
\delta^{(3)}({\bm k}+{\bm q}) \, .
\end{equation}
However, from (\ref{super_initial}), we have already seen that
the super-horizon mode $\chi_k$ is independent of $k$, and thus
 can be pulled out of the integral. Hence, we only have to
integrate over the relevant super-horizon scale momentum, for
which the upper limit is $k=k_c$.\footnote{Note,
however, that mathematically there seems no apparent reason to
 set the upper limit of the integral at $k=k_c$.
It seems reasonable to extend the range of integration into
sub-horizon scales up to an arbitrary ultraviolet cutoff
at $k=k_\mathrm{UV}$ with $k_\mathrm{UV}\gg k_c$.
If proceeding with the sub-horizon mode function
solution (\ref{sub_initial}), one finds that the squared
mode function $|\chi_k|^2$ is suppressed by a factor of
$k_c/k$ relative to super-horizon modes.
But this suppression factor is not strong enough to make
the integral independent of the ultraviolet cutoff.
Since the integrand
$|\chi_p|^2\left|\chi_{|{\bm k}-{\bm p}|}\right|^2$
is proportional to $p^{-2}$, the integral will be
dominated by the contribution from the ultraviolet cutoff,
leading to the result in proportional to $k_{\mathrm UV}$.
Of course there is a natural choice for the cutoff in the present case;
$k_{\mathrm UV}=\alpha k_c$, up to which the modes become tachyonic,
as advocated in Ref.~\cite{Lyth:2010ch,Lyth:2010zq}. If we are to take this choice,
then the resulting amplitude of curvature perturbations will be
substantially enhanced, though the qualitative result will not
change.
Nevertheless, this strong dependence of super-horizon fluctuations
on the ultraviolet cutoff deep inside the horizon looks physically
strange because it seems to imply the violation of causality.
 In fact if this were indeed the case, then we would have a first example in which
 the $\delta{N}$ formalism fails even for the curvature perturbation on super-horizon scales.
This may be originated from our assumption of the knowledge of the
{\em entire} universe beyond the horizon scale in the Fourier transformation.
 We discuss this point of maintaining causality regarding the horizon scale
 patches in the inflating universe in Appendix~\ref{app:subhorizon}, justifying
 (\ref{deltapsisquare}) and (\ref{mode_integration}) which are the very
 foundation of our computation of the correlation functions.
}
Therefore, using (\ref{super_initial}), we finally
obtain
\begin{equation}
\left\langle \left( \delta\chi^2 \right)_{\bm k}
\left( \delta\chi^2 \right)_{\bm q} \right\rangle
= (2\pi)^3 \delta^{(3)}({\bm k}+{\bm q}) \frac{4}{3^{11/3}
\left[\Gamma(2/3)\right]^4} \alpha^{-4/3} \frac{H_0^4}{k_c^3} \, .
\end{equation}

Since this expression has, as it should, the correct delta
function dependence, we can readily extract the power spectrum
$\calP_{\cal R}$. Noting (\ref{Rtochi2}) we find
\begin{equation}\label{spectrum}
\calP_{\cal R} \equiv\frac{k^3}{2\pi^2}P_{\cal R}
= \frac{32\pi^2}{3^{11/3}\left[\Gamma(2/3)\right]^4}
 \frac{\alpha^{-22/3}}{n_f} \left( \frac{k}{k_c} \right)^3 \, ,
\end{equation}
where the numerical coefficient reads
$32\pi^2/\left\{3^{11/3}\left[\Gamma(2/3)\right]^{4}\right\} \approx 1.67255$.
Thus, with $n_f =\mathcal{O}(1)$, the maximum amplitude is found at
$k=k_c$ as $\calP_{\cal R} \sim \alpha^{-22/3}$ which is already much smaller
than unity for $\alpha\gg1$.
 For larger scales, it is exponentially
suppressed and thus becomes absolutely negligible: for example,
for a scale that exited the horizon at 50 $e$-folds before waterfall,
it is suppressed by a factor $(e^{-50})^3 \approx 10^{-65}$.
As already discussed in the previous section, setting
 $\calP_{\cal R} \propto k^{n_{\cal R}-1}$, the spectrum is very blue
with the index $n_{\cal R}=4$.

\subsection{Bispectrum}

Having found the curvature perturbation and the solution of the mode function,
it is now straightforward to calculate the three-point correlation function.
We can start from the definition
\begin{align}
\left\langle \calR({\bm k}_1)\calR({\bm k}_2)\calR({\bm k}_3) \right\rangle
= & (2\pi)^3 \delta^{(3)}({\bm k}_1+{\bm k}_2+{\bm k}_3)
 B_{\cal R}({\bm k}_1,{\bm k}_2,{\bm k}_3)
\nonumber\\
= & \left( \frac{-1}{2\alpha n_f^{1/2}}
\frac{1}{\left\langle \delta\chi_S^2 \right\rangle} \right)^3
 \left\langle \left( \delta\chi^2 \right)_{{\bm k}_1}
\left( \delta\chi^2 \right)_{{\bm k}_2}
\left( \delta\chi^2 \right)_{{\bm k}_3} \right\rangle
\nonumber\\
= & \left( \frac{-1}{2\alpha n_f^{1/2}}
\frac{1}{\left\langle \delta\chi_S^2 \right\rangle} \right)^3
\int \frac{d^3q_1d^3q_2d^3q_3}{(2\pi)^{3\cdot3}}
\nonumber\\
& \hspace{3.6cm} \times \left\langle
\left( \delta\chi_{{\bm q}_1}\delta\chi_{{\bm k}_1-{\bm q}_1} \right)
\left( \delta\chi_{{\bm q}_2}\delta\chi_{{\bm k}_2-{\bm q}_2} \right)
\left( \delta\chi_{{\bm q}_3}\delta\chi_{{\bm k}_3-{\bm q}_3} \right)
\right\rangle \, .
\end{align}

As before, we are interested in the connected graphs.
This means we only take contractions between those coming
from different $(\delta\chi)_{\bm k}$'s. It is immediately seen
that there are 8 possible contractions:
for one of the two $\delta\chi_k$'s in $\calR({\bm k}_1)$,
there are four choices of contractions to one of $\delta\chi_k$'s
in $\calR({\bm k}_2)$ and $\calR({\bm k}_3)$,
and for the remaining $\delta\chi_k$ in $\calR({\bm k}_1)$,
there are two ways of contraction to either $\calR({\bm k}_2)$ or
$\calR({\bm k}_3)$ which are not chosen by the first contraction.
This gives the total number of $4\times2=8$ different contractions.
These are explicitly written as
\begin{align}
&
\contraction {\langle(} {\delta\chi}
 {{}_{{\bm q}_1}\delta\chi_{{\bm k}_1-{\bm q}_1})(} {\delta\chi}
\contraction[2ex] {\delta\chi_{{\bm q}_1}} {\delta\chi_{{\bm k}_1-}}
{{}_{{\bm q}_1})(\delta\chi_{{\bm q}_2}\delta\chi_{{\bm k}_2-{\bm q}_2})(\delta}
 {\chi_{{\bm q}_3}}
\contraction{(\delta\chi_{{\bm q}_1}\delta\chi_{{\bm k}_1-{\bm q}_1})(\delta\chi_{{\bm q}_2}}
 {\delta\chi_{{\bm k}}}
 {{}_{{}_2-{\bm q}_2})(\delta\chi_{{\bm q}_3}} {\delta\chi_{|}}
\langle
(\delta\chi_{{\bm q}_1}\delta\chi_{{\bm k}_1-{\bm q}_1})
 (\delta\chi_{{\bm q}_2}\delta\chi_{{\bm k}_2-{\bm q}_2})
 (\delta\chi_{{\bm q}_3}\delta\chi_{{\bm k}_3-{\bm q}_3})
\rangle +
\contraction {\langle(} {\delta\chi} {{}_{{\bm q}_1}\delta\chi_{{\bm k}_1-{\bm q}_1})(}
 {\delta\chi}
\contraction[2ex] {\delta\chi_{{\bm q}_1}}
{\delta\chi_{{\bm k}_1-}} {{}_{{\bm q}_1})
(\delta\chi_{{\bm q}_2}\delta\chi_{{\bm k}_2-{\bm q}_2})(\delta\chi_{{\bm q}_3}}
{\delta\chi_{{\bm k}_3-}}
\contraction {(\delta\chi_{{\bm q}_1}\delta\chi_{{\bm k}_1-{\bm q}_1})(\delta\chi_{{\bm q}_2}}
 {\delta\chi_{{\bm k}}} {{}_{{}_2-{\bm q}_2})(} {\delta\chi_{{\bm q}_3}}
\langle
(\delta\chi_{{\bm q}_1}\delta\chi_{{\bm k}_1-{\bm q}_1})
 (\delta\chi_{{\bm q}_2}\delta\chi_{{\bm k}_2-{\bm q}_2})
 (\delta\chi_{{\bm q}_3}\delta\chi_{{\bm k}_3-{\bm q}_3})
\rangle
\nonumber\\
+ &
\contraction {(} {\delta\chi_{{\bm q}_1}}
 {\delta\chi_{{\bm k}_1-{\bm q}_1})(\delta\chi_{{\bm q}_2}} {\delta\chi_{{\bm k}}}
\contraction[2ex] {\delta\chi_{{\bm q}_1}} {\delta\chi_{{\bm k}_1-}}
 {{}_{{\bm q}_1})(\delta\chi_{{\bm q}_2}\delta\chi_{{\bm k}_2-{\bm q}_2})(\delta}
 {\chi_{{\bm q}}}
\contraction[3ex] {(\delta\chi_{{\bm q}_1}\delta\chi_{{\bm k}_1-{\bm q}_1})(}
 {\delta\chi_{{\bm q}_2}} {\delta\chi_{{\bm k}_2-{\bm q}_2})(\delta\chi_{{\bm q}_3}}
 {\delta\chi_{{\bm k}}}
\langle
(\delta\chi_{{\bm q}_1}\delta\chi_{{\bm k}_1-{\bm q}_1})
 (\delta\chi_{{\bm q}_2}\delta\chi_{{\bm k}_2-{\bm q}_2})
 (\delta\chi_{{\bm q}_3}\delta\chi_{{\bm k}_3-{\bm q}_3})
\rangle +
\contraction {(} {\delta\chi_{{\bm q}_1}}
 {\delta\chi_{{\bm k}_1-{\bm q}_1})(\delta\chi_{{\bm q}_2}} {\delta\chi_{{\bm k}}}
\contraction[2ex] {\delta\chi_{{\bm q}_1}} {\delta\chi_{{\bm k}_1-}}
 {{}_{{\bm q}_1})(\delta\chi_{{\bm q}_2}\delta\chi_{{\bm k}_2-{\bm q}_2})
(\delta\chi_{{\bm q}_3}} {\delta\chi_{{\bm k}_3-}}
\contraction[3ex] {(\delta\chi_{{\bm q}_1}\delta\chi_{{\bm k}_1-{\bm q}_1})(}
 {\delta\chi_{{\bm q}_2}} {\delta\chi_{{\bm k}_2-{\bm q}_2})(} {\delta\chi_{{\bm q}}}
\langle
(\delta\chi_{{\bm q}_1}\delta\chi_{{\bm k}_1-{\bm q}_1})
 (\delta\chi_{{\bm q}_2}\delta\chi_{{\bm k}_2-{\bm q}_2})
 (\delta\chi_{{\bm q}_3}\delta\chi_{{\bm k}_3-{\bm q}_3})
\rangle
\nonumber\\
+ &
\contraction {(} {\delta\chi_{{\bm q}_1}} {\delta\chi_{{\bm k}_1-{\bm q}_1})
(\delta\chi_{{\bm q}_2}\delta\chi_{{\bm k}_2-{\bm q}_2})(} {\delta\chi_{{\bm q}}}
\contraction[2ex] {\delta\chi_{{\bm q}_1}}
 {\delta\chi_{{\bm k}_1-}}{{}_{{\bm q}_1})(\delta} {\chi_{{\bm q}_2}}
\contraction[2ex] {(\delta\chi_{{\bm q}_1}\delta\chi_{{\bm k}_1-{\bm q}_1})
(\delta\chi_{{\bm q}_2}} {\delta\chi_{{\bm k}}} {{}_{{}_2-{\bm q}_2})
(\delta\chi_{{\bm q}_3}} {\delta\chi_{{\bm k}}}
\langle
(\delta\chi_{{\bm q}_1}\delta\chi_{{\bm k}_1-{\bm q}_1})
 (\delta\chi_{{\bm q}_2}\delta\chi_{{\bm k}_2-{\bm q}_2})
 (\delta\chi_{{\bm q}_3}\delta\chi_{{\bm k}_3-{\bm q}_3})
\rangle +
\contraction {(} {\delta\chi_{{\bm q}_1}}
 {\delta\chi_{{\bm k}_1-{\bm q}_1})(\delta\chi_{{\bm q}_2}\delta\chi_{{\bm k}_2-{\bm q}_2})(}
 {\delta\chi_{{\bm q}}}
\contraction[2ex] {\delta\chi_{{\bm q}_1}}
 {\delta\chi_{{\bm k}_1-}} {{}_{{\bm q}_1})
(\delta\chi_{{\bm q}_2}\delta} {\chi_{{\bm k}_2}}
\contraction[3ex] {(\delta\chi_{{\bm q}_1}\delta\chi_{{\bm k}_1-{\bm q}_1})(}
 {\delta\chi_{{\bm q}_2}} {\delta\chi_{{\bm k}_2-{\bm q}_2})(\delta\chi_{{\bm q}_3}}
 {\delta\chi_{{\bm k}}}
\langle
(\delta\chi_{{\bm q}_1}\delta\chi_{{\bm k}_1-{\bm q}_1})
 (\delta\chi_{{\bm q}_2}\delta\chi_{{\bm k}_2-{\bm q}_2})
 (\delta\chi_{{\bm q}_3}\delta\chi_{{\bm k}_3-{\bm q}_3})
\rangle
\nonumber\\
+ &
\contraction {(} {\delta\chi_{{\bm q}_1}}
{\delta\chi_{{\bm k}_1-{\bm q}_1})(\delta\chi_{{\bm q}_2}
\delta\chi_{{\bm k}_2-{\bm q}_2})(\delta\chi_{{\bm q}_3}} {\delta\chi_{{\bm k}}}
\contraction[2ex] {\delta\chi_{{\bm q}_1}}
 {\delta\chi_{{\bm k}_1-}} {{}_{{\bm q}_1})(\delta} {\chi_{{\bm q}}}
\contraction[2ex] {(\delta\chi_{{\bm q}_1}
\delta\chi_{{\bm k}_1-{\bm q}_1})(\delta\chi_{{\bm q}_2}}
{\delta\chi_{{\bm k}}} {{}_{{}_2-{\bm q}_2})(}{\delta\chi_{{\bm q}}}
\langle
(\delta\chi_{{\bm q}_1}\delta\chi_{{\bm k}_1-{\bm q}_1})
 (\delta\chi_{{\bm q}_2}\delta\chi_{{\bm k}_2-{\bm q}_2})
 (\delta\chi_{{\bm q}_3}\delta\chi_{{\bm k}_3-{\bm q}_3})
\rangle +
\contraction {(} {\delta\chi_{{\bm q}_1}}
 {\delta\chi_{{\bm k}_1-{\bm q}_1})(\delta\chi_{{\bm q}_2}
\delta\chi_{{\bm k}_2-{\bm q}_2})(\delta\chi_{{\bm q}_3}} {\delta\chi_{{\bm k}}}
\contraction[2ex] {\delta\chi_{{\bm q}_1}}
 {\delta\chi_{{\bm k}_1-}} {{}_{{\bm q}_1})(\delta\chi_{{\bm q}_2}\delta}
 {\chi_{{\bm k}_2}}
\contraction[3ex] {(\delta\chi_{{\bm q}_1}\delta\chi_{{\bm k}_1-{\bm q}_1})(}
 {\delta\chi_{{\bm q}_2}} {\delta\chi_{{\bm k}_2-{\bm q}_2})(}{\delta\chi_{{\bm q}}}
\langle
(\delta\chi_{{\bm q}_1}\delta\chi_{{\bm k}_1-{\bm q}_1})
 (\delta\chi_{{\bm q}_2}\delta\chi_{{\bm k}_2-{\bm q}_2})
 (\delta\chi_{{\bm q}_3}\delta\chi_{{\bm k}_3-{\bm q}_3})
\rangle \, .
\end{align}
Each of these terms exactly corresponds to the term
with $\delta^{(3)}({\bm k}_1+{\bm k}_2+{\bm k}_3)$.
Indeed, we find these 8 contractions give
\begin{equation}
8\delta^{(3)}({\bm k}_1+{\bm k}_2+{\bm k}_3)
\int d^3q \left| \chi_q \right|^2
\left| \chi_{|{\bm k}_1-{\bm q}|} \right|^2
\left| \chi_{|{\bm k}_2+{\bm q}|} \right|^2 \, .
\end{equation}
Again noting that $|\chi_k|^2$ is independent of momentum, and has a
 cut-off at $k=k_c$, we obtain
\begin{align}
& \int \frac{d^3q_1d^3q_2d^3q_3}{(2\pi)^{3\cdot3}}
\left\langle \left( \delta\chi_{{\bm q}_1}\delta\chi_{{\bm k}_1-{\bm q}_1} \right)
 \left( \delta\chi_{{\bm q}_2}\delta\chi_{{\bm k}_2-{\bm q}_2} \right)
 \left( \delta\chi_{{\bm q}_3}\delta\chi_{{\bm k}_3-{\bm q}_3} \right) \right\rangle
\nonumber\\
& = 8\delta^{(3)}({\bm k}_1+{\bm k}_2+{\bm k}_3) \frac{4\pi k_c^3}{3}
 \left[ \frac{2\sqrt{\pi}}{3^{2/3}\Gamma(2/3)}
\frac{H_0}{\sqrt{2k_c^3}\alpha^{1/3}} \right]^6 \, .
\end{align}
Comparing this expression with the definition of the bispectrum,
 we find
\begin{equation}
B_{\cal R}({\bm k}_1,{\bm k}_2,{\bm k}_3)
= -\frac{16(2\pi)^7}{3^5\left[\Gamma(2/3)\right]^6}
\frac{\alpha^{-11}}{n_f^{3/2}k_c^6} \, ,
\end{equation}
where the numerical coefficient reads
$16(2\pi)^7/\left\{3^{5}[\Gamma(2/3)]^{6}\right\} \approx 4128.89$.
 To leading order, the bispectrum has no momentum dependence,
and thus the dimensionless shape function
$(k_1k_2k_3)^2B_{\cal R}({\bm k}_1,{\bm k}_2,{\bm k}_3)$ exhibits
 its maximum amplitude at the equilateral limit $k_1=k_2=k_3$.
This is anticipated, since the curvature perturbation produced by
the waterfall field is {\em intrinsically} highly non-Gaussian.
Note, however, that this bispectrum is completely unobservable on
large scales: in the equilateral limit, multiplying $k^6$, we see
for example that it is exponentially suppressed
 by a factor of $(e^{-50})^6 \approx 10^{-130}$ for a scale that
exited the horizon at 50 $e$-folds before the waterfall.
Thus this bispectrum is totally hopeless to be detected
on large scales.

\section{Distribution of curvature perturbation}
\label{sec:distribution}

In this section, we consider the one-point probability distribution
function of $\calR$.
Basically, we can guess the form of the probability distribution function.
At leading order $\calR$ is proportional to the square of
$\delta\chi_L$ which is very close to Gaussian.
Thus, the probability distribution of $\calR \sim \delta\chi_L^2$
is expected to be very close to the chi-squared distribution.

The fully nonlinear distribution function can be obtained from
(\ref{deltapsisquare}).
By setting $n=n_f+\delta n$ and the left hand side of it to be $m^2/g^2$,
and regarding $\delta n=\calR$ as a function of $\delta\chi_L$,
the distribution function $\mathbb{P}$ of $\calR$ is given as
\begin{equation}\label{Rpdf}
\mathbb{P}(\calR)
 = \mathbb{P}_{\chi}(\delta\chi_L)\frac{d\delta\chi_L}{d\calR} \, .
\end{equation}
Here, we already know that $\mathbb{P}_{\chi}(\delta\chi_L)$
 is a Gaussian distribution with zero mean, i.e.
\begin{equation}
\mathbb{P}_\chi(\delta\chi_L)
 = \frac{1}{\sqrt{2\pi}\sigma_{\delta\chi_L}}
\exp \left( -\frac{\delta\chi_L^2}{2\sigma_{\delta\chi_L}^2} \right) \, ,
\end{equation}
and the variance
$\sigma_{\delta\chi_L}^2\equiv\langle\delta\chi_L^2\rangle-\langle\delta\chi_L\rangle^2$,
 with $\langle{\delta\chi_L}\rangle=0$, is given by
\begin{equation}
\sigma_{\delta\chi_L}^2 = \left\langle \delta\chi_L^2 \right\rangle
 = \int d\log{k} \calP_{\delta\chi_L}(k)\, ,
\end{equation}
where the power spectrum $\calP_{\delta\chi_L}(k)$ of the
fluctuations $\delta\chi_L$ can be found
from (\ref{super_initial}) as
\begin{equation}\label{chifluc_spectrum}
\calP_{\delta\chi_L}(k)=\frac{k^3}{2\pi^2}|\delta\chi_k|^2
 = \frac{\alpha^{-2/3}H_0^2}{3^{4/3}\pi
 \left[\Gamma(2/3)\right]^2} \left( \frac{k}{k_c} \right)^3 \,.
\end{equation}
Noting that the large scale modes has a cut-off
at $k=k_c$, we obtain
\begin{equation}\label{variance_x}
\sigma_{\delta\chi_L}^2 = \frac{\alpha^{-2/3}H_0^2}{3^{7/3}\pi
\left[\Gamma(2/3)\right]^2} \, .
\end{equation}

\begin{figure}[htb]
 \begin{center}
  \includegraphics[width=10cm]{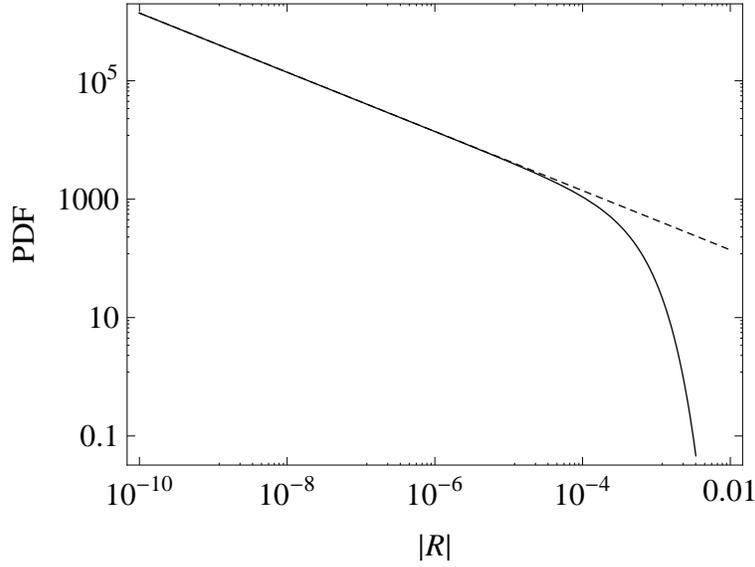}\vspace{-1.5em}
 \end{center}
 \caption{The probability distribution of $|\calR|$ (\ref{PDF}), with $r \approx 0.01$ and $\alpha \approx 8$.
As $|\calR|$ becomes larger, the probability drops down extremely sharply.
 For comparison, we also show the chi-squared distribution (dotted line) with appropriate normalization.
 At small value of $|\calR|$, the two distribution functions behave in the
 same manner but they become different at larger $|\calR|$.}
 \label{fig:PDF}
\end{figure}

Now, we evaluate (\ref{deltapsisquare}) at $n = n_f+\delta n=n_f+\calR$
to write
\begin{equation}
\frac{m^2}{g^2} =  \left( \delta\chi_L^2 + \frac{\alpha^2H_0^2}{8\pi^2} \right) A^2
 \exp \left[ \frac{4}{3}\alpha(n_f+\calR)^{3/2} - 3(n_f+\calR) \right] \, .
\end{equation}
This equation can be easily solved for $\delta\chi_L$ as a function of $\calR$,
\begin{equation}\label{xintermsofy}
\delta\chi_L = \sqrt{ \frac{m^2}{A^2g^2}
\exp \left[ -\frac{4}{3}\alpha(n_f+\calR)^{3/2} + 3(n_f+\calR) \right]
- \frac{\alpha^2H_0^2}{8\pi^2} } \, .
\end{equation}
Thus from (\ref{Rpdf}) we can immediately find the probability distribution of $\calR$ as
\begin{align}\label{PDF}
\mathbb{P}(\calR)
= & \frac{1}{\sqrt{2\pi}\sigma_{\delta\chi_L}}
\exp \left\{ -\frac{1}{2\sigma_{\delta\chi_L}^2}
\left[ \frac{m^2}{A^2g^2} e^{-\frac{4}{3}\alpha(n_f+\calR)^{3/2} + 3(n_f+\calR)}
 - \frac{\alpha^2H_0^2}{8\pi^2} \right] \right\}
\nonumber\\
& \times \left\{ \frac{m^2}{A^2g^2}
\exp \left[ -\frac{4}{3}\alpha(n_f+\calR)^{3/2} + 3(n_f+\calR) \right]
 - \frac{\alpha^2H_0^2}{8\pi^2} \right\}^{-1/2}
\left( \alpha\sqrt{n_f+\calR} -\frac{3}{2} \right)
\nonumber\\
& \times \frac{m^2}{A^2g^2}
\exp \left[ -\frac{4}{3}\alpha(n_f+\calR)^{3/2} + 3(n_f+\calR) \right] \, .
\end{align}
This is a fairly complex probability distribution function, and is very
 different from the Gaussian one. We plot it in Fig.~\ref{fig:PDF}.

Having the distribution function of $\calR$ at hand,
let us consider the mean value $\langle{\calR}\rangle$.
We can formally write it as
\begin{equation}
\langle{\calR}\rangle = \int \calR\mathbb{P}(\calR)d\calR
= \int \calR(\delta\chi_L)\mathbb{P}(\delta\chi_L)d\delta\chi_L \, ,
\end{equation}
where $\calR$ is now regarded as a function of $\delta\chi_L$.
Although we cannot invert (\ref{xintermsofy}) to find $\calR(\delta\chi_L)$
exactly, we can obtain an approximate expression by assuming
$|\calR|\ll 1$ as
\begin{equation}
\calR = \left\{ \frac{3}{4\alpha} \log \left[ \frac{m^2}{g^2}
\left( \delta\chi_L^2 + \frac{\alpha^2H_0^2}{8\pi^2} \right)^{-1} \right] \right\}^{2/3}
 - n_f \approx -\frac{(2\pi)^2}{\alpha^3n_f^{1/2}}
 \left( \frac{\delta\chi_L}{H_0} \right)^2 + \cdots \, ,
\end{equation}
where we have expanded in the limit
$\delta\chi_L^2\ll\alpha^2H_0^2/(8\pi^2)$. It is trivial to find that
for a Gaussian distribution $\mathbb{P}(x)$,
\begin{equation}
\int x^2\mathbb{P}(x)dx = \sigma_x^2 \, ,
\end{equation}
and thus the average value of $\calR$ is found, using (\ref{variance_x}),
as
\begin{equation}
\langle\calR\rangle
= -\frac{4\pi}{3^{7/3}\left[\Gamma(2/3)\right]^2}
\frac{\alpha^{-11/3}}{n_f^{1/2}} \, .
\end{equation}
The overall numerical factor is
 $4\pi/\left\{3^{7/3}\left[\Gamma(2/3)\right]^{2}\right\} \approx 0.527976$.
If we take $n_f = \mathcal{O}(1)$, the most important factor is
its dependence on $\alpha$: for $\alpha\gg1$, it is indeed very small.
If we could have $\alpha$ of order unity, the mean value could become
large. But as we have discussed in Sec.~\ref{evolution},
 this cannot be the case
because of the condition for an efficient tachyonic instability.

As we have mentioned before, the relevant curvature perturbation is
not $\calR$ itself but $\calR-\langle\calR\rangle$.
Therefore, although $\calR$ is always negative, the true
fluctuations from the mean value can become positive.
Nevertheless, since the mean value $\langle\calR\rangle$ turns out to
be very small, there is no chance to have a large positive fluctuation:
the fluctuation is bounded from above as
\begin{eqnarray}
\calR-\langle\calR\rangle\leq|\langle\calR\rangle|\,.
\end{eqnarray}
Recalling that $\calR$ is negative of the gravitational
 potential, we can see that the curvature perturbation induced by the
 waterfall field {\em repels} matter around rather than attract.
This implies that there would be no primordial black hole formation
even on scales as small as the Hubble horizon scale,
but rather bubbles of void may appear. But this is a highly qualitative
 argument and we need more explicit calculations,
 which we do not pursue in this note.

\section{Conclusion}
\label{sec:conclusion}

In this paper, we have examined the contribution of the waterfall field
 $\chi$ to the curvature perturbation. The waterfall field $\chi$ can
change the final curvature perturbation during the period between the
moment of waterfall and the end of inflation, because $\chi$ controls
the physical processes during this time: the waterfall phase transition
occurs as soon as the effective mass squared of $\chi$ becomes negative,
and the end of inflation is determined by the mean square fluctuations
of the sub-horizon modes of $\chi$ which became tachyonic after the
waterfall transition.

By solving the equation of $\chi$, we have obtained for both the super- and
 sub-horizon modes the amplitudes at the moment of waterfall and time
dependence until the end of inflation in terms of the number of $e$-folds.
 Using the $\delta{N}$ formalism, we have calculated both the power
spectrum and bispectrum of the curvature perturbation induced by the
waterfall field $\chi$. The power spectrum is steeply blue with $n_\calR=4$,
 and the bispectrum exhibits the maximum amplitude at the equilateral limit.
 This indicates that the distribution of the curvature perturbation is
intrinsically non-Gaussian, and we have presented the explicit form of
the distribution function. On large scales, however, both the power
spectrum and bispectrum are exponentially suppressed and totally negligible.

\subsection*{Acknowledgement}

We thank Hassan Firouzjahi and David Wands for fruitful discussions.
We are especially indebted to David Lyth for sharing Ref.~\cite{Lyth:2010zq}
before it appeared on arXiv and many important correspondences
regarding the contributions of the sub-horizon modes.
We are also grateful to Viatcheslav Mukhanov for
his warm hospitality at Arnold Sommerfeld Center for Theoretical Physics, LMU in Munich
where part of this work was carried out.
JG appreciates Ana Ach\'ucarro for helpful conversations that motivated this work.
JG is grateful to the Yukawa Institute for Theoretical Physics, Kyoto University
 for hospitality during the long-term workshop ``Gravity and Cosmology 2010 (GC2010)''
 (YITP-T-10-01) and the YKIS symposium ``Cosmology -- The Next Generation --'' (YKIS2010),
where this work was initiated, and the 20th Workshop on General Relativity and Gravitation in Japan (YITP-W-10-10) where this work was being finished.
This work was supported in part
by a VIDI and a VICI Innovative
Research Incentive Grant from the Netherlands Organisation for Scientific Research (NWO),
Korea Institute for Advanced Study under the KIAS Scholar program,
by the Grant-in-Aid for the Global COE Program at Kyoto University,
``The Next Generation of Physics, Spun from Universality and Emergence''
from the Ministry of Education, Culture, Sports, Science and Technology (MEXT) of Japan,
by JSPS Grant-in-Aid for Scientific Research (A) No.~21244033,
by JSPS Grant-in-Aid for Creative Scientific Research No.~19GS0219,
and by Alexander von Humboldt Foundation.

\appendix

\section{Evaluation of $\calR$ from linear perturbation equation}
\label{app:conventional}

Here to check if our result based on the $\delta N$ formalism
is consistent with the standard perturbation theory, we evaluate
the curvature perturbation by using the linear perturbation
equation for $\calR$. See also Ref.~\cite{Lyth:2010ch} for this approach.

In linear theory, on super-horizon scales,
it is known that the curvature perturbation
on comoving slices $\calR$ satisfies
\begin{eqnarray}
\dot\calR=-H\frac{\delta P_{c}}{\rho+P}\,,
\label{dotcalR}
\end{eqnarray}
where $\delta P_c$ is the pressure perturbation
on comoving slices. The comoving slice is defined
by $\delta T^0{}_i=0$. In the present case, this means
\begin{eqnarray}
\delta T^0{}_i=-\left(\dot\phi\partial_i\delta\phi
+\dot\chi\partial_i\chi\right)
\approx-\phi\partial_i\delta\phi=0\,,
\end{eqnarray}
where we have used the fact that
$\chi=\delta\chi_L$ and $\calP_{\delta\chi_L}(k)\propto k^3$,
namely the fact that  on super-horizon scales
the contribution from the waterfall field
to $\delta T^0{}_i$ is negligible
compared to that from the inflaton field. That is, on super-horizon scales,
the comoving slices are defined solely in terms of the inflaton
as those on which the inflaton field is homogeneous.

Therefore the contribution to the pressure perturbation
$\delta P_c$ comes totally from the waterfall field $\chi$,
\begin{eqnarray}
\delta P_c=\frac{1}{2}H_0^2{\delta\chi_L'}^2
+\frac{1}{2}(M^2-g^2\phi^2)\delta\chi_L^2\,.
\label{deltaPc}
\end{eqnarray}
As for $\rho+P$, we have
\begin{eqnarray}
\rho+P
=H_0^2\left[\phi'{}^2+\langle\delta\chi_S'{}^2\rangle\right]\,.
\end{eqnarray}

Let us evaluate $\phi'{}^2$ and
$\langle\delta\chi_S'{}^2\rangle$ to see which term dominates
during waterfall.
For $\phi'{}^2$ we have
\begin{eqnarray}
\phi'{}^2(n)=(r\phi_c)^2e^{-2rn}
=H_0^2\frac{r^2\beta}{g^2}e^{-2rn}\,.
\end{eqnarray}
For $\langle\delta\chi_S'{}^2\rangle$ we have at $n\gtrsim1$
\begin{eqnarray}\label{deltachiprime}
\langle\delta\chi_S'{}^2(n)\rangle
\approx
\alpha^2n\langle\delta\chi_S^2(n)\rangle
\approx
H_0^2\frac{\alpha^4}{8\pi^2}A^2
\exp \left( \frac{4}{3}\alpha n^{3/2} - 3n\right) \,.
\end{eqnarray}
Hence the time dependent ratio of $\langle\delta\chi_S'{}^2\rangle$ to ${\phi'}^2$ is written as
\begin{eqnarray}
R(n)\equiv\frac{\langle\delta\chi_S'{}^2\rangle}{\phi'{}^2}
\approx\frac{2g^2\alpha^3A^2}{8\pi^2r}
\exp\left(\frac{4}{3}\alpha n^{3/2} - 3n\right)\,.
\label{ratio1}
\end{eqnarray}
Using (\ref{endofinf}) and (\ref{deltachiprime}), at the end of inflation we have
\begin{eqnarray}
R(n_f)=\frac{\langle\delta\chi_S'{}^2(n_f)\rangle}{\phi'{}^2(n_f)}
\approx6n_f\,.
\label{ratiofin}
\end{eqnarray}
Therefore, for $n_f\gtrsim 1$,
 $\langle\delta\chi_S'{}^2\rangle$ becomes dominant toward the
end of inflation.
Using this result, we can rewrite (\ref{ratio1}) as
\begin{eqnarray}\label{Rn}
R(n)=R(n_f)\frac{R(n)}{R(n_f)}
\approx6n_f\exp\left[\frac{4}{3}\alpha (n^{3/2}-n_f^{3/2}) - 3(n-n_f)\right]\,.
\end{eqnarray}
Let $n_\mathrm{eq}\equiv n_f-\Delta n$ be the time at which $\langle\delta\chi_S'{}^2\rangle$
begins to dominate over $\phi'{}^2$. Since the growth rate of
$\langle\delta\chi_S'{}^2\rangle$ is very fast and
the ratio $R(n_f)$ at the end of inflation (\ref{ratiofin}) is not so large,
$\sim10$ or so, the $\langle\delta\chi_S'{}^2\rangle$-dominated stage appears
only at the very near the end of inflation, $\Delta n\ll1$. Specifically,
setting $R(n_\mathrm{eq})=1$, we find
\begin{eqnarray}
\Delta n\approx \frac{\ln(6n_f)}{2\alpha n_f^{1/2}}\sim \frac{1}{2\alpha n_f^{1/2}}\,.
\label{Deltan}
\end{eqnarray}
Therefore, $\phi'{}^2$, which is almost constant in time, dominates
over $\langle\delta\chi_S'{}^2\rangle$ almost all the stage of the waterfall
$n\lesssim n_\mathrm{eq}$.

With the above result in mind, we rewrite (\ref{dotcalR}) as
\begin{eqnarray}
\frac{d\calR}{dn}&=&-\frac{\delta P_{c}}{\rho+P}
\approx
-\frac{\delta\chi_L^2(n)}{\langle\delta\chi_S^2(n)\rangle}\,
\frac{R(n)}{1+R(n)} \, .
\label{dcalRdn}
\end{eqnarray}
Since $\delta\chi_L^2(n)/{\langle\delta\chi_S^2(n)\rangle}$
is time-independent, we can just replace it by that evaluated at
$n=0$.
Then (\ref{dcalRdn}) can be expressed as
\begin{eqnarray}
\frac{d\calR}{dn}&=&
-\frac{\delta\chi_L^2(0)}{\langle\delta\chi_S^2(0)\rangle}\,
\frac{R(n)}{1+R(n)}\,.
\label{dcalRdn2}
\end{eqnarray}
The last factor on the right hand side is negligible
for $n<n_\mathrm{eq}$ and approximately equal to one for $n_\mathrm{eq}<n<n_f$.
Therefore, with the initial condition that $\calR(0)=0$,
it can be easily integrated to give
\begin{eqnarray}
\calR(n_f)\approx
-\frac{\delta\chi_L^2(0)}{\langle\delta\chi_S^2(0)\rangle}\Delta n\,.
\label{finalR}
\end{eqnarray}
With the identification that $\Delta n\approx 1/(2\alpha n_f^{1/2})$
as evaluated in (\ref{Deltan}),
this agrees with our result using the $\delta N$ formalism\footnote{
Note that if we faithfully integrate (\ref{dcalRdn2}) using (\ref{Rn}), we can even recover
the logarithmic correction factor $\ln(6n_f)$ in (\ref{Deltan}) as the leading order
approximation of the integral.
}.

\section{Short wavelength modes}
\label{app:subhorizon}

In this section, we justify (\ref{deltapsisquare}) and the integration of (\ref{mode_integration}),
and argue why we do not go beyond the horizon scale.
We consider a scalar field $\phi(\bx)$ and decompose
it into Fourier modes $\widetilde{\phi}(\bk)$.
Let us call the modes with wavelengths smaller than the
horizon size $L_H=2\pi/H$ the short wavelength modes and those
larger than $L_H$ the long wavelength modes. We assume
the universe is inflating.

When we decompose $\phi(\bx)$, usually we assume  we have
the knowledge of the whole (infinitely large) universe.
That is,
\begin{eqnarray}
\phi(\bx)=\int \frac{d^3k}{(2\pi)^3}\,\widetilde\phi(\bk)\,e^{i\bk\cdot\bx}
\quad
\leftrightarrow
\quad
\widetilde\phi(\bk)=\int d^3x\,\phi(\bx)\,e^{-i\bk\cdot\bx}\,.
\end{eqnarray}
If we divide the above into those composed of
long wavelength modes and short wavelength modes,
\begin{eqnarray}
\phi(\bx)
=\phi_L(\bx)+\phi_S(\bx)
=\int_{k<H} \frac{d^3k}{(2\pi)^3}\,\widetilde\phi(\bk)e^{i\bk\cdot\bx}
+ \int_{k>H} \frac{d^3k}{(2\pi)^3}\,\widetilde\phi(\bk)e^{i\bk\cdot\bx}\,,
\label{whole}
\end{eqnarray}
then this will naturally induce a non-zero correlation between
$\phi_S(\bx)$ and $\phi_S(\by)$ even if the two points
are separated at a distance larger than the horizon size,
\begin{eqnarray}
\left\langle\phi_S(\bx)\phi_S(\by)\right\rangle\neq0
\end{eqnarray}
for $|\bx-\by|> L_H$.
Since each horizon size region should be causally unrelated
during inflation, this result is {\em acausal}.
This is apparently due to our
assumption that we, i.e. the observers belonging to different
regions of horizon size, have the knowledge of
the whole universe.

Therefore, in stead of (\ref{whole}),
it is more reasonable to divide the field in such a
way that $\phi_S(\bx)$ and $\phi_S(\by)$ will not be correlated
if $|\bx-\by|> L_H$.
To incorporate this prescription, we proceed as follows.
We introduce two boxes of different
size, a very large box $L^3$ where $L=NL_H$ with $N\gg1$ being a very large integer and
the horizon size box $L_H^3$.
The large box would correspond to the present horizon size
of the universe.

We define $\chi_S(\bx)$ for each horizon size box as
\begin{eqnarray}
\phi_{S(i)}(\bx)=\theta_{(i)}(\bx)
\sum_{{\bk}}\widetilde\phi_{(i)}(\bk)e^{i\bk\cdot(\bx-\bx_i)}
\quad \left(\bk=\frac{2\pi}{L_H}{\bm n}\right)\,,
\end{eqnarray}
where ${\bm n}=(n_1,n_2,n_3)$ ($n_i$ are integers), $\bx_i$ is
the center of $i$-th box, and
$\theta_{(i)}(\bx)=1$ if $\bx$ is in the $i$-th horizon size region
and zero otherwise.
The long wavelength part is defined by
\begin{eqnarray}
\phi_L(\bx)=\sum_{|{\bm n}|\leq N}\widetilde\phi_L(\bk)e^{i\bk\cdot\bx}
\quad \left(\bk=\frac{2\pi}{L}{\bm n}
=\frac{2\pi}{L_H}\frac{\bm n}{N}\right)\,.
\end{eqnarray}
Thus we have the decomposition,
\begin{eqnarray}
\phi(\bx) & = & \phi_L(\bx)+\phi_S(\bx)\,,
\\
\phi_S(\bx) & = & \sum_{i}\phi_{S(i)}(\bx)\,.
\end{eqnarray}
This guarantees that there is no correlation of between
two short wavelength modes that belong to
two different horizon size regions: for $|\bx-\by|> L_H$,
\begin{eqnarray}
\left\langle\phi_S(\bx)\phi_S(\by)\right\rangle=0 \, .
\end{eqnarray}

Now we take the square of $\phi(\bx)$ and average over
the horizon scale. We obtain
\begin{eqnarray}
\left\langle\phi^2(\bx)\right\rangle_{L_H}
&=&\phi_L^2(\bx)+\left\langle\phi_S^2(\bx)\right\rangle
\cr
&=&\phi_L^2(\bx)+\sum_{i}\left\langle\phi_{S(i)}^2(\bx)\right\rangle
=\phi_L^2(\bx)
+\sum_{i}\theta_{(i)}(\bx)\sum_{\bk}\left|\widetilde\phi_{(i)}(\bk)\right|^2
\,.
\end{eqnarray}
It is reasonable to assume that $|\widetilde\phi_{(i)}(\bk)|^2$ is
independent of the region $(i)$. Hence we may set
$\left|\widetilde\phi_{(i)}(\bk)\right|^2=\left|\widetilde\phi_S(\bk)\right|^2$.
Then since $\sum_i\theta_{(i)}(\bx)=1$,
we obtain
\begin{eqnarray}
\left\langle\phi^2(\bx)\right\rangle_{L_H}
=\phi_L^2(\bx)
+\sum_{i}\theta_{(i)}(\bx)\sum_{\bk}\left|\widetilde\phi_{S}(\bk)\right|^2
=\phi_L^2(\bx)
+\sum_{\bk}\left|\widetilde\phi_{S}(\bk)\right|^2\,.
\end{eqnarray}
This agrees with (\ref{deltapsisquare}).

Also, if we consider the sum on the short wavelength modes,
\begin{eqnarray}
\sum_{\bp}\left|\widetilde\phi_{(i)}(\bp)\right|^2\left|\widetilde\phi_{(i)}(\bp+\bk)\right|^2
=\sum_{\bp}\left|\widetilde\phi_{S}(\bp)\right|^2\left|\widetilde\phi_{S}(\bp+\bk)\right|^2\,,
\end{eqnarray}
which appears in (\ref{mode_integration}), it is apparent
that this is non-vanishing only
for $|\bk|\geq2\pi/L_H$, because there exists no sum for
$|\bk|<2\pi/L_H$ by definition. This means there will be
no contribution from the short wavelength modes to the
curvature perturbation on super-horizon scales.

\end{document}